\documentclass[sigconf]{acmart}
\usepackage{balance}
\usepackage{bm}
\usepackage{adjustbox}
\usepackage{graphicx,xcolor} 
\usepackage[flushleft]{threeparttable}
\usepackage{balance}
\usepackage{algorithm}
\usepackage{algpseudocode}
\usepackage{multirow}

\usepackage{listings}
\usepackage{xcolor}
\usepackage{amsmath}

\renewcommand{\Comment}[2][.4\linewidth]{%
  \leavevmode\hfill\makebox[#1][l]{//~#2}}

\definecolor{codegreen}{rgb}{0,0.6,0}
\definecolor{codegray}{rgb}{0.5,0.5,0.5}
\definecolor{codepurple}{rgb}{0.58,0,0.82}
\definecolor{backcolour}{rgb}{0.95,0.95,0.92}

\lstdefinestyle{mystyle}{
  backgroundcolor=\color{backcolour}, commentstyle=\color{codegreen},
  keywordstyle=\color{magenta},
  numberstyle=\tiny\color{codegray},
  stringstyle=\color{codepurple},
  basicstyle=\ttfamily\footnotesize,
  breakatwhitespace=false,         
  breaklines=true,                 
  captionpos=b,                    
  keepspaces=true,                 
  numbers=left,                    
  numbersep=5pt,                  
  showspaces=false,                
  showstringspaces=false,
  showtabs=false,                  
  tabsize=2
}
\algblock{ParFor}{EndParFor}
\algnewcommand\algorithmicparfor{\textbf{for}}
\algnewcommand\algorithmicpardo{\textbf{parallel do}}
\algnewcommand\algorithmicendparfor{\textbf{end\ parallel for}}
\algrenewtext{ParFor}[1]{\algorithmicparfor\ #1\ \algorithmicpardo}
\algrenewtext{EndParFor}{\algorithmicendparfor}

\AtBeginDocument{%
  \providecommand\BibTeX{{%
    \normalfont B\kern-0.5em{\scshape i\kern-0.25em b}\kern-0.8em\TeX}}}

\copyrightyear{2022}
\acmYear{2022}
\setcopyright{rightsretained}
\acmConference[FPGA '22]{Proceedings of the 2022 ACM/SIGDA International Symposium on Field-Programmable Gate Arrays}{February 27-March 1, 2022}{Virtual Event, CA, USA}
\acmBooktitle{Proceedings of the 2022 ACM/SIGDA International Symposium on Field-Programmable Gate Arrays (FPGA '22), February 27-March 1, 2022, Virtual Event, CA, USA}
\acmDOI{10.1145/3490422.3502359}
\acmISBN{978-1-4503-9149-8/22/02}

\begin{document}
\fancyhead{}
\title{HP-GNN: Generating High Throughput \underline{GNN} Training Implementation on CPU-FPGA \underline{H}eterogeneous \underline{P}latform}

\author{Yi-Chien Lin}
\authornote{Both authors contributed equally to this research.}
\email{yichienl@usc.edu}
\affiliation{%
  \institution{University of Southern California}
  \city{Los Angeles}
  \state{California}
  \country{USA}
}

\author{Bingyi Zhang}
\email{bingyizh@usc.edu}
\authornotemark[1]
\affiliation{%
  \institution{University of Southern California}
  \city{Los Angeles}
  \state{California}
  \country{USA}
}

\author{Viktor Prasanna}
\email{prasanna@usc.edu}
\affiliation{%
  \institution{University of Southern California}
  \city{Los Angeles}
  \state{California}
  \country{USA}
}

\renewcommand{\shortauthors}{Lin and Zhang, et al.}

\begin{abstract}
  Graph Neural Networks (GNNs) have shown great success in many applications such as recommendation systems, molecular property prediction, traffic prediction, etc.
  Recently, CPU-FPGA heterogeneous platforms have been used to accelerate many applications by exploiting customizable data path and  abundant user-controllable on-chip memory resources of FPGAs. 
  Yet, accelerating and deploying GNN training on such platforms requires not only
  expertise in hardware design but also substantial development efforts.

  We propose HP-GNN, a novel framework that generates high throughput GNN training implementations on a given CPU-FPGA platform that can benefit both application developers and machine learning researchers.
  HP-GNN takes GNN training algorithms, GNN models as the inputs, and automatically performs hardware mapping onto the target CPU-FPGA platform. 
HP-GNN consists of: (1) data layout and internal representation that reduce the memory traffic and random memory accesses; (2) optimized hardware templates that support various GNN models; (3) a design space exploration engine for automatic hardware mapping; (4) high-level application programming interfaces (APIs) that allows users to specify GNN training with only a handful of lines of code.
  To evaluate HP-GNN, we experiment with two well-known sampling-based GNN training algorithms and two GNN models. For each training algorithm and model, HP-GNN generates implementation on a state-of-the-art CPU-FPGA platform. 
  Compared with CPU-only and CPU-GPU platforms,  experimental results show that the generated implementations achieve $55.67\times$ and $2.17\times$ speedup on the average, respectively. Compared with the state-of-the-art GNN training implementations, HP-GNN achieves up to $4.45\times$ speedup.
  
  
\end{abstract}



\maketitle

\section{Introduction}
Recently, Graph Neural Networks (GNNs) have shown great success in many fields including recommendation systems \cite{ying2018graph, zhu2019aligraph}, molecular property prediction \cite{hamilton2017inductive}, traffic prediction \cite{jiang2021graph}, etc.
Initially, GNN was trained using the full graph \cite{gcn} as the input, directly. However, as graphs become larger, full-graph GNN training becomes inefficient because the model is only updated once for each epoch. Moreover, huge amount of intermediate results needs to be stored in full-graph GNN training, which leads to high memory footprint \cite{sample_survey}. To overcome the above issues, many sampling-based GNN training algorithms \cite{hamilton2017inductive, graphsaint-iclr20, chen2018fastgcn, chiang2019cluster} have been proposed for training the GNN models. Sampling-based methods first sample the full graph to produce mini-batches, and then take these mini-batches as input for training.
The sampling-based mini-batch training methods demonstrate great  advantages compared with full-graph training  in  terms  of accuracy,  generalization,  and  scalability for large-scale graphs \cite{hamilton2017inductive, graphsaint-iclr20, chen2018fastgcn, chiang2019cluster}. Therefore, state-of-the-art GNN frameworks \cite{Fey/Lenssen/2019, wang2019dgl} adopt  sampling-based mini-batch training algorithms.

Compared with the CPU-only platforms or CPU-GPU platforms, CPU-FPGA platforms are promising platforms to deploy GNN training since this can support customized memory access pattern \cite{zhou2019hitgraph} and data layout to reduce the substantial memory traffic and random memory access in GNN training.
However, deploying GNN training on CPU-FPGA heterogeneous platform is challenging due to notorious development efforts that requires hardware design expertise.
Though there are many FPGA-based GNN accelerators \cite{hygcn, zhang2021boostgcn, zhang2020hardware, geng2020awb, linGCN}, previous works either focus on full-graph GNN inference, or a specific GNN model, or a specific GNN training algorithm, and no general framework has been proposed. 

Motivated by the challenges, we propose HP-GNN, a framework for mapping  GNN training on CPU-FPGA heterogeneous platform. 
We first formulate an high-level abstraction to describe the computation of sampling-based mini-batch GNN training. Then we develop optimized  hardware templates based on the GNN abstraction.
In order to reduce development effort and eliminate the need for hardware expertise, HP-GNN provides easy-to-use software programming APIs that allow fast development without the need for hardware programming.
 To achieve high throughput and automate the accelerator generation process, we develop a general design space exploration engine that optimizes the hardware configuration based on the selected GNN training algorithm. In addition, we propose a data layout and internal representation that reduces the memory traffic and random memory access.  We summarize our contributions as follow:



\begin{itemize}
    \item We propose a general framework for mapping various sampling based mini-batch GNN training onto a CPU-FPGA platform.  We demonstrate the applicability of HP-GNN to various sampling algorithms and GNN models.
    \item We provide high-level and easy-to-use software programming interface that abstracts the hardware implementation details. 
    \item To enable hardware mapping  and  high throughput GNN training, the proposed framework consists of the following optimizations:
    \begin{itemize}
        \item Data layout and internal representation that reduce the memory traffic and random memory access caused by the irregular computation pattern of GNN. 
        \item Optimized hardware templates that support various widely-used GNN models.
        \item  General design space exploration algorithm that generates hardware design configuration to optimize the training throughput for various sampling methods and algorithmic parameters. 
    \end{itemize}
    \item We evaluate our framework using two state-of-the-art GNN models: GraphSAGE \cite{hamilton2017inductive}, GCN \cite{gcn}, along with two commonly used sampling methods: neighbor sampling \cite{hamilton2017inductive}, and subgraph sampling \cite{graphsaint-iclr20}. Running on a Xilinx Alveo U250 board hosted by a 64-core AMD processor, the experimental results show that the accelerators produced by our framework can achieve $2.17\times$ throughput on the average compared with a state-of-the-art GNN framework on CPU-GPU platform.
\end{itemize}


\section{Background and related work}

\subsection{GNN Models}
\label{GNN_train}
\begin{figure}[h]
    \centering
    \includegraphics[width=7cm]{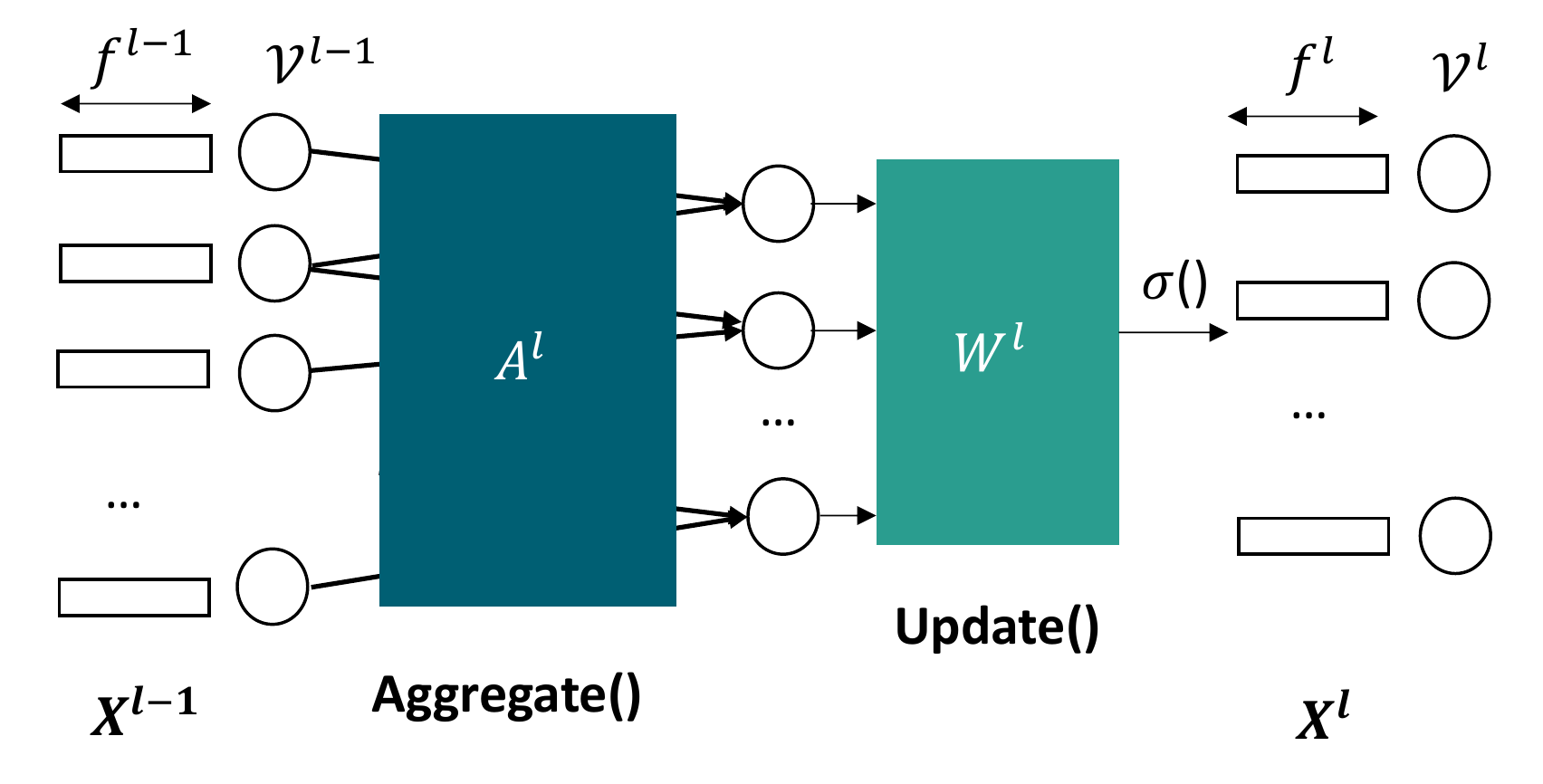}
    \vspace{-0.3cm}
    \caption{Computation abstraction of a  GNN layer}
     \label{fig:high-level}
     \vspace{-0.15cm}
\end{figure} 
Given an input graph $\mathcal{G}(\mathcal{V}, \mathcal{E}, \bm{X})$,  a GNN model is specified by:
\begin{itemize}
    \item $L$: number of layers.
    \item $\mathcal{V}^{t}$: a set of target vertices to be inferred.
    \item $f^{l}$: hidden dimension in layer $l~( 1 \leqslant l \leqslant L)$.
    \item A mechanism of how to construct: 
    \begin{itemize}
        \item $\mathcal{V}^{l}$: the set of vertices in layer $l~( 0 \leqslant l \leqslant L) $. $|\mathcal{V}^{l}|$ denotes the number of vertices in layer $l$. Moreover, $\mathcal{V}^{L} = \mathcal{V}^{t}$.
         \item $\bm{A}^{l} \in \mathbb{R}^{{|\mathcal{V}^{l-1}|}\times {|\mathcal{V}^{l}|}} $: adjacency matrix for feature aggregation in layer $l~( 1 \leqslant l \leqslant L)$. $\bm{A}^{l}$ defines the inter-layer connectivity between $\mathcal{V}^{l-1}$ and $\mathcal{V}^{l}$. 
    \end{itemize}

    \item $\bm{W}^{l}\in \mathbb{R}^{f^{l-1}\times f^{l}}$: weight matrix of layer $l~( 1 \leqslant l \leqslant L)$ that is used in update function to perform linear transformation of vertex features.
    \item $\bm{X}^{l}\in \mathbb{R}^{{|\mathcal{V}^{l}|}\times f^{l}}$: feature matrix of layer $l~( 1 \leqslant l \leqslant L)$
    \item Aggregate() function that is used by each vertex to aggregate information from its neighbors.
    \item Update() function including an one-layer multi-layer perception (MLP) and an activation function $\sigma$() that is used by each vertex to perform feature transformation.
\end{itemize}

Figure \ref{fig:high-level} depicts the computation abstraction of a GNN layer. The computations using a GNN model can also be expressed using the aggregate-update paradigm \cite{hygcn}, as shown in Algorithm \ref{alg:aggregate-update-paradigm-define}. $\bm{h}_{v}^l \in \mathbb{R}^{f^{l}}$ is the feature vector of $v\in \mathcal{B}^{l}$ in layer $l$, and $\bm{a}_{v}^l \in \mathbb{R}^{f^{l}} $ is the intermediate result of $v\in \mathcal{B}^{l}$.
\vspace{-0.1cm}
\begin{algorithm}
\caption{GNN Computation Abstraction}
\label{alg:aggregate-update-paradigm-define}

\begin{algorithmic}[1]
\For{$l=1...L$}
\For{vertex $v \in \mathcal{V}^l$}
\State{$\bm{a}^l_{v} = \textbf{Aggregate(}\bm{h}_{u}^{l-1}: u\in \mathcal{N}(v)$ and $u\in \mathcal{V}^{l-1}\textbf{)}$}
\State{$\bm{h}_{v}^l = \textbf{Update(}\bm{a}_i^l, \bm{W}^{l}, \sigma() \textbf{)}$}
\EndFor
\EndFor
\end{algorithmic}
\end{algorithm}
\vspace{-0.2cm}
There are several widely used GNN models proposed in the literature:

\textbf{GCN}: GCN is proposed in \cite{gcn}. Given the input graph $\mathcal{G}(\mathcal{V}, \mathcal{E}, \bm{X})$, the GCN model is specified by:
\begin{itemize}
    \item $\mathcal{V}^{1} =\mathcal{V}^{2} = ...=\mathcal{V}^{L} = \mathcal{V}^{T} = \mathcal{V}$
    \item $\bm{A}^{1} =\bm{A}^{2} = ...= \bm{A}^{L} = \bm{D}^{-\frac{1}{2}}(\bm{A}+\bm{I})\bm{D}^{-\frac{1}{2}} $, where $\bm{A}$ and $\bm{D}$ are the adjacency matrix and the Laplacian matrix of the input graph. $\bm{I}$ is the identity matrix.
    \item $L$: number of layers; $f^{l}$: feature size in layer $l~( 1 \leqslant l \leqslant L)$; 
\end{itemize}
 The Aggregate() function and Update() function of GCN are expressed as: 
\begin{equation}
    \begin{split}
        \bm{a}_{v}^{l} & = \text{Sum}(\frac{1}{ \sqrt{D(v)\cdot D(u)}} \cdot \bm{h}_{u}^{l-1}:u\in \mathcal{N}(v)\cup  \{v\} )\\
        \bm{h}_{v}^{l} & = \text{ReLU} \left(\bm{a}_{v}^{l}\bm{W}^{l} + \bm{b}^{l} \right)
    \end{split}
\end{equation}
where $\mathcal{N}(v)$ denotes the neighbor set of $v$ in $\mathcal{V}^{l-1}$, $D(v)$ denotes the degree of vertex $v$, and $\bm{b}^{l}$ denotes the bias of the update function.

\textbf{GraphSAGE}: GraphSAGE is proposed in \cite{hamilton2017inductive} for inductive representation learning on graphs. Starting from a set of target vertex $\mathcal{V}^{T}$, GraphSAGE neighbor sampler recursively samples the neighbors to build $\mathcal{V}^{1}, \mathcal{V}^{2}, ...,\mathcal{V}^{L} $. The adjacency matrix $\bm{A}^{l}$ defines the edge connections between $\mathcal{V}^{l-1}$ and $\mathcal{V}^{l}$, and each  edge has  weight equal to one. The Aggregate() function and Update() function of GraphSAGE are expressed as: 
\begin{equation}
    \begin{split}
        \bm{a}_{v}^{l} & = \bm{h}_{v}^{l-1} || \text{Mean} \left(  \bm{h}_{u}^{l-1}:u\in\mathcal{N}(v) \cup \{v\} \right) \\
        \bm{h}_{v}^{l} & = \text{ReLU}  \left(\bm{a}_{v}^{l}\bm{W}^{l} + \bm{b}^{l} \right)
    \end{split}
\end{equation}

\textbf{Note}: In the rest of the paper, we use GCN and GraphSAGE to refer to their GNN-layer operators Aggregate(), Update().

\subsection{GNN training}
To train a GNN model, the GNN training process consists of five stages \cite{hamilton2017inductive, graphsaint-iclr20, chen2018fastgcn, chiang2019cluster}: sampling, forward propagation, loss calculation, back propagation and weight update. In the sampling stage, a set of vertices and adjacency matrices are sampled from $\{\mathcal{V}^{l}:0\leqslant l \leqslant L\}$ and $\{\bm{A}^{l}:1\leqslant l \leqslant L\}$ to form a mini-batch. We use $\mathcal{B}^l$ to denote the vertices sampled from $\mathcal{V}^l$ in layer $l$. $\bm{A}_{s}^l$ denotes the sampled adjacency matrix, which describes inter-layer connections (edges) between $\mathcal{B}^l$ and $\mathcal{B}^{l-1}$ within the mini-batch. A mini-batch consists of target vertices $\mathcal{B}^L$, sampled vertices for each layer $\{\mathcal{B}^{l}:0\leqslant l\leqslant L-1\}$, and sampled adjacency matrices $\{\bm{A}_{s}^1:1\leqslant l\leqslant L-1\}$. In the forward propagation stage, the mini-batch is processed layer by layer.  The output embeddings in the last layer $\{\bm{h}_{i}^{L}:v_{i}\in \mathcal{B}^{L}\}$ are compared with the ground truth for loss calculation. The calculated loss will be used as input for back propagation, which performs a similar computation as forward propagation but in the reverse direction. Finally, the gradients of $\bm{W}^l$ in each layer are derived and be used to update the weight. We show the steps of GNN training in Algorithm \ref{alg:GNN}. In Algorithm \ref{alg:GNN}, $\mathcal{N}_{s}(v)$ denotes neighbors of $v$ in $\mathcal{B}^{l-1}$ that are specified in $\bm{A}_{s}^{l}$. {A GNN training algorithm is specified by} a sampling algorithm (see Section \ref{sampling}) to construct the mini-batch that consists of  $\{\mathcal{B}^{l}:0\leqslant l\leqslant L-1\}$ and $\{\bm{A}_{s}^1:1\leqslant l\leqslant L-1\}$.

In HP-GNN, we exploit data parallelism within each mini-batch by aggregating and updating multiple vertices concurrently (shown in Line 5 of Algorithm \ref{alg:GNN}). The computation order within the same mini-batch does not affect the final results. Thus, training in our parallel framework leads to the same result and accuracy as training in serial fashion.
\begin{small}
\begin{algorithm}
\caption{Sampling-based GNN Training Algorithm}
\label{alg:GNN}

\begin{algorithmic}[1]
\For{each iteration}
\State{\textbf{Sampling( )}} 
{\color{blue}\Comment{Derive mini-batches}}
\For{$l=1...L$}
{\color{blue}\Comment{Forward Propagation}}
\For{vertex $v \in \mathcal{B}^l$}
\State{$\bm{a}^l_{v} = \textbf{Aggregate(}\bm{h}_{u}^{l-1}: u\in \mathcal{N}_{s}(v)$ and $u\in \mathcal{B}^{l-1}\textbf{)}$}
\State{$\bm{h}_{v}^l = \textbf{Update(}\bm{a}_i^l, \bm{W}^{l}, \sigma() \textbf{)}$}
\EndFor
\EndFor
\State{\textbf{CalculateLoss($\{\bm{h}_{i}^{L}:v_{i}\in \mathcal{B}^{L}\}$)}}
\State{\textbf{BackPropagation( )}}
{\color{blue}\Comment{Derive gradient of ${W}^l$}}
\State{\textbf{WeightUpdate( )}}

\EndFor

\end{algorithmic}
\end{algorithm}
\end{small}


\vspace{-0.2cm}
\subsection{Sampling Algorithm}\label{sampling}
{An algorithm to sample a mini-batch is specified by}:
\begin{itemize}
    \item A method to sample the vertices $\mathcal{B}^{l}~(0 \leqslant l \leqslant L)$ from $\mathcal{V}^{l}$.
    \item A method to construct the adjacency matrix $\bm{A}_{s}^{l}~(1 \leqslant l \leqslant L)$ from $\bm{A}^{l}$. 
\end{itemize}
By sampling the vertices $\mathcal{B}^{l}~(0 \leqslant l \leqslant L)$ and the adjacency matrix $\bm{A}^{l}~(1 \leqslant l \leqslant L)$, we construct the mini-batch as the input for each training iteration. 
Various sampling methods \cite{graphsaint-iclr20, hamilton2017inductive, chen2018fastgcn, ying2018graph, dai2018learning} are proposed to form a mini-batch from input graphs. These sampling methods \cite{sample_survey} falls into three categories: neighbor sampling, layer-wise sampling and subgraph sampling. We only introduce neighbor sampling and subgraph sampling, since layer-wise sampling \cite{chen2018fastgcn} has the similar computation pattern with  subgraph sampling \citep{graphsaint-iclr20}.

\vspace{0.1cm}
 \noindent \textbf{Neighbor Sampling:}
Neighbor sampling \cite{hamilton2017inductive, ying2018graph, chen2017stochastic}  is a type of sampling strategy that recursively samples the neighbors from the target vertices. To perform neighbor sampling, users specify the size of target vertices $|\mathcal{V}^t|$ and the neighbor sample size $NS^l$ for each vertex in layer $l$.
The sampler first chooses a set of vertices as target vertices $\mathcal{V}^t$. Then, the sampler samples $NS^L$ neighbors for each vertex in $\mathcal{V}^t$ based on a specific probability distribution (e.g., uniform distribution \cite{hamilton2017inductive}). After the first iteration of neighbor sampling, the set of 1-hop sampled neighbors $\mathcal{B}^{L-1}$ is obtained, where $|\mathcal{B}^{L-1}|=|\mathcal{V}^t|\times NS^L$. Similarly, we obtain the 2-hop neighbors $\mathcal{B}^{L-2}$ by sampling the neighbors of the 1-hop neighbors $\mathcal{B}^{L-1}$, where $|\mathcal{B}^{L-2}|=|\mathcal{V}^t|\times NS^L\times NS^{L-1}$.  By performing neighbor sampling recursively, we obtain $L$-hop neighbors of the target vertices. After obtaining  $\{\mathcal{B}^{l}:0\leqslant l\leqslant L-1\}$ ,  $\bm{A}_{s}^{l}~(1\leqslant l \leqslant L)$ is constructed by:
\begin{equation}
    \bm{A}_{s}^{l}(u,v) =
    \begin{cases}
         \bm{A}^{l}(u,v) & u\in \mathcal{B}^{l-1} \text{ and } v\in \mathcal{B}^{l} \\
         0 & \text{otherwise}
    \end{cases}
\end{equation}

\vspace{0.1cm}
\noindent\textbf{Subgraph Sampling:}
Subgraph sampling \cite{graphsaint-iclr20, chiang2019cluster} is a strategy to sample a subgraph from the input graph and perform GNN information propagation within the subgraph.
In the subgraph sampling-based method, users specify sampling budget $SB$ which denotes how many vertices to be sampled for the subgraph. Then, the sampler samples $SB$ of vertices or edges based on specific probability, and induce a subgraph based on the sampled vertices or edges.  For subgraph sampling, the sampled vertices are identical for each layer, i.e. $\mathcal{B}^0=\mathcal{B}^1=...=\mathcal{B}^L$. The construction of  $\bm{A}_{s}^{l}~(1\leqslant l \leqslant L)$ is the same as neighbor sampling.


\subsection{Related work}

\subsubsection{Software GNN Frameworks}
Several software GNN frameworks \citep{Fey/Lenssen/2019, wang2019dgl, zhu2019aligraph, wang2021gnnadvisor} have been proposed in the literature. PyTorch Geometric (PyG) \cite{Fey/Lenssen/2019} is one of the most commonly-used frameworks for GNN deployment that uses PyTorch \cite{NEURIPS2019_9015} as the backend. PyG users can describe various GNN models using PyG's message passing API. Deep Graph Library (DGL) \cite{wang2019dgl} adopts several parallelization strategies to achieve high performance and memory efficiency for GNN computations. Moreover, DGL offers several major frameworks \cite{NEURIPS2019_9015,199317,DBLP:journals/corr/ChenLLLWWXXZZ15} as the backend, allowing users to port their model across frameworks.
Aligraph \cite{zhu2019aligraph} supports training on heterogeneous attributed graph (HAG), i.e. graphs that contain different types of vertices and edges. 
These software frameworks share similar features: They are built on existing frameworks \cite{NEURIPS2019_9015,199317,DBLP:journals/corr/ChenLLLWWXXZZ15}, and abstract away the detailed implementations by providing graph-oriented APIs so that users can describe GNN models easily.

\begin{figure*}[h]
    \centering
    \includegraphics[width=13.5cm]{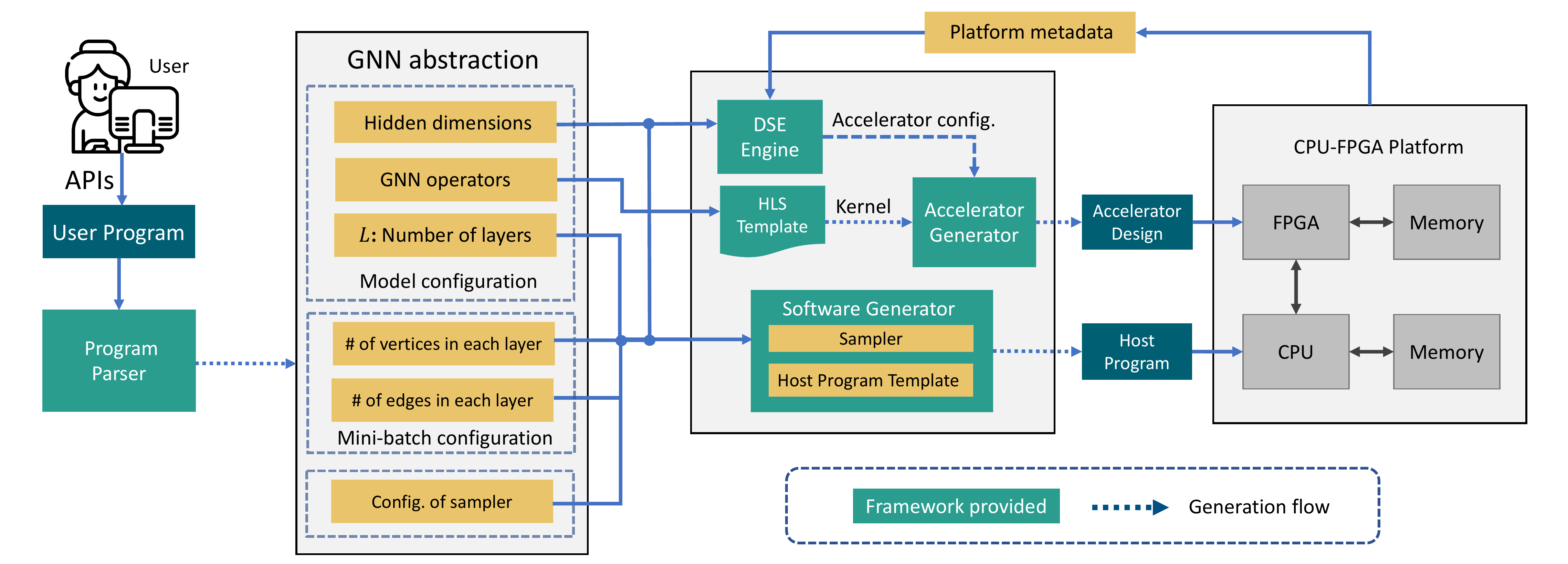}
    \vspace{-0.3cm}
    \caption{Framework overview}
     \label{fig:framework}
     \vspace{-0.3cm}
\end{figure*}

\subsubsection{Hardware GNN Acceleration}
GraphACT \cite{10.1145/3373087.3375312} performs sub-graph sampling based GNN training on CPU-FPGA heterogeneous platform. GraphACT exploits both task-level parallelism and data parallelism, and adopts a redundancy reduction technique to reduce the number of on-chip memory accesses. However, GraphACT optimizes the hardware design for inductive GNN models \cite{hamilton2017inductive} using subgraph sampling, and does not support transductive model, such as GCN . As shown in \cite{hu2021open}, different sampling algorithms perform well in different applications, so there is no one-size-fits-all sampling algorithms. 
Rubik \cite{9428002} decomposes GNN computations into graph-level and node-level computations, and proposes hierarchical task mapping strategies to exploit data reuse and parallelism in the two computation levels. However, Rubik is an ASIC accelerator that is hard to be optimized for various sampling algorithms.
DeepBuring-GL \cite{9256539} is a FPGA framework to accelerate GNN inference. DeepBuring-GL provides various templates to support different GNN computations and memory access patterns.  DeepBuring-GL analyzes the GNN model with the input graph to identify performance bottleneck and choose appropriate hardware templates for kernel implementation to accelerate GNN inference.
 Though frameworks like DeepBuring-GL have been proposed, most of the hardware acceleration works still require significant hardware expertise to make use of them. Moreover, no framework has been proposed to support various mini-batch GNN training on CPU-FPGA platform, which motivates us to conduct this work. In this paper, we build a framework to accelerate GNN training on CPU-FPGA platform. Our framework provides infrastructures to support various GNN models and training algorithms.
 
 We summarize the benefits of using CPU-FPGA platform: while CPU can support various sampling algorithms, FPGA platform allows customizable data path and memory access pattern that can be exploited to optimize the GNN training throughput.

\section{Framework}
\subsection{System Overview}
\label{architecture}

\begin{figure}[h]
    \centering
    \includegraphics[width=7.5cm]{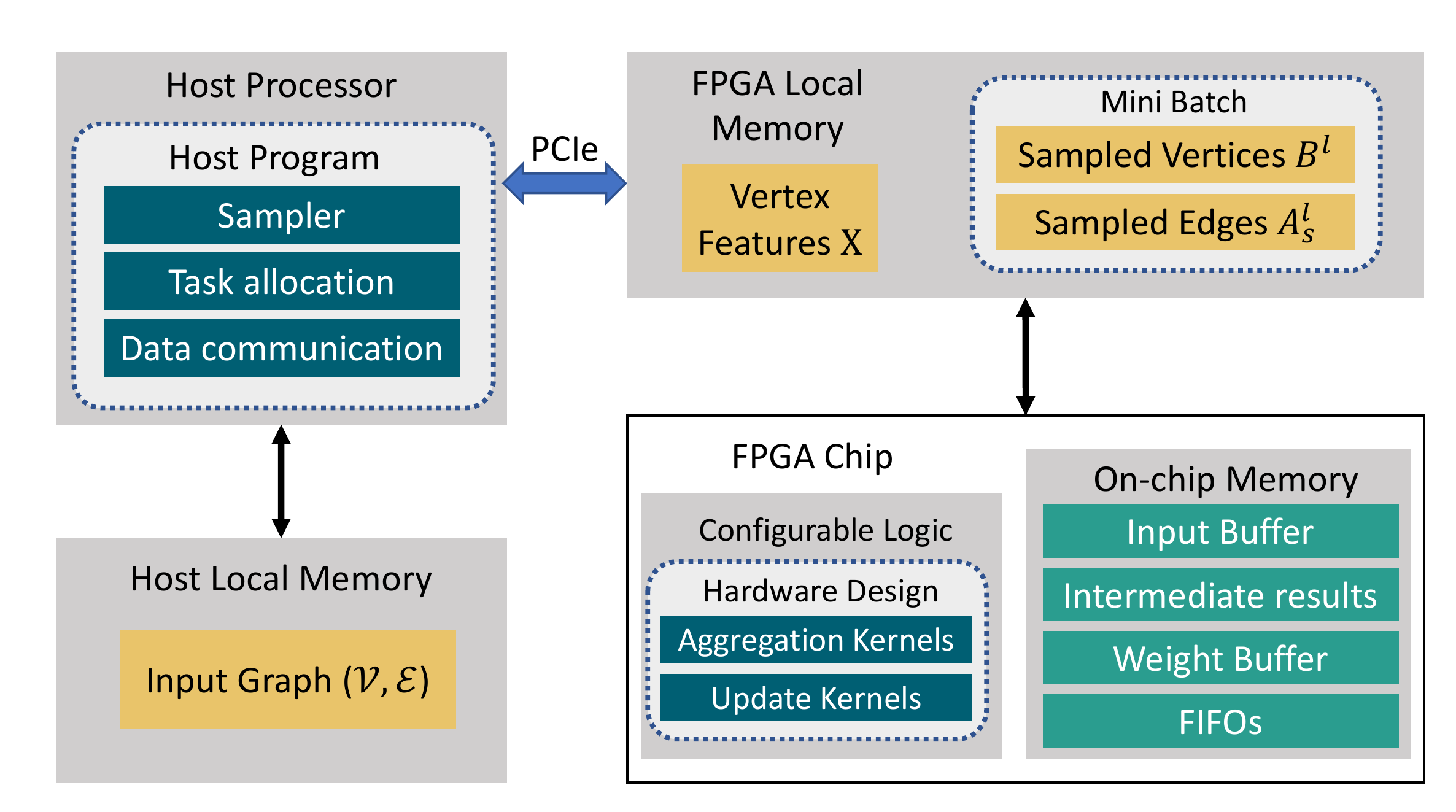}
   \vspace{-0.3cm}
    \caption{System overview}
     \label{fig:arch}
     \vspace{-0.3cm}
\end{figure}

Figure \ref{fig:arch} depicts the mapping of graph data and various kernels onto the CPU-FPGA heterogeneous platform. Sampling is performed on the host CPU because CPU is flexible to support various sampling algorithms; GNN operations including feature aggregation and feature update are performed on the  proposed FPGA accelerator. 
Based on this task assignment, the structural information of input graph ($\mathcal{V}, \mathcal{E}$) is stored in the host memory for the host CPU to perform sampling. After sampling is done, the structural information of the mini-batch is generated and transferred to the FPGA local memory. The vertex features $\bm{X}$ are stored in the FPGA local memory to be directly accessed by the FPGA accelerator, which can reduce the overhead of data movement. The state-of-the-art FPGA boards \cite{fpga260} have up to 260 GB memory; this can support medium size graph. 

Regarding very large graphs, we store the vertex features in host memory and transfer the vertex features of the mini-batch to the FPGA accelerator after sampling.

\subsection{Framework Overview}
\label{framework}
Figure \ref{fig:framework} demonstrates the framework overview. The generated design by the framework consists of two major components: (1) a host program that manages CPU-FPGA communication, kernel task scheduling and mini-batch sampling; (2) an accelerator design that runs on the FPGA. 
To generate the mini-batch GNN training implementation on CPU-FPGA platform, our framework takes the user program as the input and generates a high-level abstraction for mini-batch GNN training. In the input program, user specifies the following parameters:
\begin{itemize}
    \item GNN parameters: number of layers $L$; hidden dimension of each layer: $f^{l}, (0\leqslant l \leqslant L)$. The hidden dimensions also define the dimension of weight matrix $\bm{W}^{l}\in \mathbb{R}^{f^{l}\times f^{l+1}}$. 
    \item Specify an off-the-shelf GNN model, or provide user-defined functions (UDFs) for scatter( ), gather( ) and update( ) to build custom GNN computation layer. 
    \item Sampling algorithm and its parameters. For example, a neighbor sampler for a 2-layer GNN model can be defined as \emph{Sampler( 'NeighborSampler', L=2, budgets=[10, 25])} through our  high-level API described in Section \ref{subsec:high-level-api}. We provide several off-the-shelf samplers for users to choose from.
\end{itemize}
The program parser extracts a GNN abstraction from user program, which serves as the intermediate representation for the software generator and hardware generator to generate the implementations on CPU-FPGA platform. The GNN abstraction consists of GNN model configuration (hidden dimensions $f^{l}$, GNN operators, number of layers $L$) and mini-batch configuration (number of vertices in each layer $|\mathcal{B}^{l}|,(0 \leqslant l \leqslant L)$ and number of edges in each layer $|\mathcal{E}^{l}|,(1 \leqslant l \leqslant L)$). The mini-batch configuration is deduced from the sampling algorithm that implies number of vertices $|\mathcal{B}^{l}|, (0 \leqslant l \leqslant L)$ in each layer and number of edges $|\mathcal{E}^{l}|, (1 \leqslant l \leqslant L)$  in each layer.



\vspace{0.1cm}
\noindent\textbf{DSE Engine}: the DSE engine takes the GNN abstraction and the platform metadata as input and generates the accelerator configuration that optimizes the GNN training throughput (Section \ref{sec:dse}). 


\vspace{0.1cm}
\noindent\textbf{HLS Template/Hardware Template}: In the framework, we provide optimized hardware templates written in high-level synthesis (HLS). The key computation operators of the templates (e.g. scatter(), gather()) are obtained from the GNN abstraction. We describe the details of the HLS template design in Section \ref{HLS_template}.

\vspace{0.1cm}
\noindent\textbf{Accelerator generator}: Given the generated accelerator configuration and hardware templates, the accelerator generator generates the hardware accelerators for the target FPGA board. The accelerator generator uses the available synthesis tools such as Xilinx Vitis as the backend.

\vspace{0.1cm}
\noindent\textbf{Software generator}: Given the input program, the software generator produces a runtime system that runs on the host processor. The runtime system performs the mini-batch sampling, CPU-FPGA communication management, and task scheduling.



\begin{table}[]
\renewcommand{\arraystretch}{1.2}
\begin{adjustbox}{max width=0.47\textwidth}
\begin{tabular}{l|l}
\toprule
\textbf{API Functions}                       &    \textbf{Description}    \\ \midrule 
\textbf{Init( )}       & Initialization the platform with FPGA bitstream                 \\  \midrule 
\textbf{GNN\_Parameters( )} & Number of layer $L$, feature length $f^l$, $W^l$ and $X$   \\ \midrule
\textbf{GNN\_Computation( )} &\begin{tabular}[|l|]{@{}l@{}}  The layer operators in GNN model\\ Specify an off-the-shelf GNN model or "customized" \end{tabular} \\ \midrule 
\textbf{Scatter( )} & UDF, required if customized layer operator is specified  \\ \midrule 
\textbf{Gather( )} &  UDF, required if customized layer operator is specified \\ \midrule 
\textbf{Update( )} &  UDF, required if customized layer operator is specified \\ \midrule 
\textbf{GNN\_Model( )} & Build GNN model using GNN parameters and computation  \\ \midrule 
\textbf{PlatformParameters( )} & FPGA Memory bandwidth, Number of DSPs, LUTs, etc.  \\ \midrule 
\textbf{LoadInputGraph( )} & Specify the input graph.  \\ \midrule 
\textbf{Sampler( )} & Sampling method with algorithmic parameters           \\ \midrule 
\textbf{DistributeData( )} & Distribute graph into host memory and FPGA local memory \\ \midrule 
\textbf{GenerateDesign( )} & Generate hardware design and software design       \\ \midrule 
\textbf{Start\_training( )} & Run GNN training     \\\midrule 
\textbf{Save\_model( )} & Save trained GNN model     \\\midrule 
\textbf{PrepareEdges( )} & Prepare graph edge values that is used for training     \\\bottomrule
\end{tabular}
\end{adjustbox}

\vspace{0.1cm}
\caption{Programming interfaces \label{tab:hostAPI}}
\vspace{-0.8cm}
\end{table}




\subsection{High-Level APIs}
\label{subsec:high-level-api}

Table \ref{tab:hostAPI} summarizes our provided high-level APIs for user to program the mini-batch training in Python. Listing \ref{lst:program} is an example for developing the GNN training using our proposed framework. 

\begin{lstlisting}[language=Python, caption=An example user program for GNN training , label={lst:program}]
# design generation phase
Samp = Sampler( 'NeighborSampler', L=2, budgets=[10,25])

GNN_comp = GNN_Computation('SAGE')
GNN_para = GNN_Parameters(L=2, hidden=[256], feat)
Model = GNN_Model(GNN_comp, GNN_para)
# specify the resoruces of a single super logic region, using Xilinx-U250 as an example
ParaFPGA = PlatformParameters(board= , SLR = 4, DSP=3072, LUT=423000, URAM=320, BRAM= , BW=19.25) 
bitstream, runtime = GenerateDesign(Sampler, Model, ParaFPGA) # generate hardware design and software design, return the pointers 

# runtime phase
Graph = LoadInputGraph('Reddit', Path='')
Graph = PrepareEdges(Graph, Model)
Init(bitstream) # initialize the hardware platform
edgepointers, featurepointers = DistributeData(Graph) # distribute graph structural information and vertex features into host memory and fpga memory 

start_training(runtime, edgepointers, featurepointers, epochs=10) # run GNN training
save_model() # save the trained model
\end{lstlisting}

In the design phase, user specifies the mini-batch sampler, GNN model and parameters of platform. The framework automatically generates the optimized accelerator design and software design. In the runtime phase, user starts the GNN training on the target CPU-FPGA platform. Using our high-level APIs, a GNN training program only requires a few dozen lines of code.

\noindent \textbf{Use cases}: Our framework can serve application developers, who utilize existing GNN models to build GNN applications. Our framework provides off-the-shelf GNN training implementations for some of the commonly-used GNN models (GCN \cite{gcn}, GraphSAGE \cite{hamilton2017inductive}, GIN \cite{xu2018powerful}) which can be directly deployed on the CPU-FPGA platform. 
For machine learning (ML) researchers who develop novel GNN models, our framework allows them to customize their own GNN models. For both cases, our framework accelerates GNN training on CPU-FPGA platform without hardware expertise.

\begin{figure*}[h]
    \centering
    \includegraphics[width=14cm]{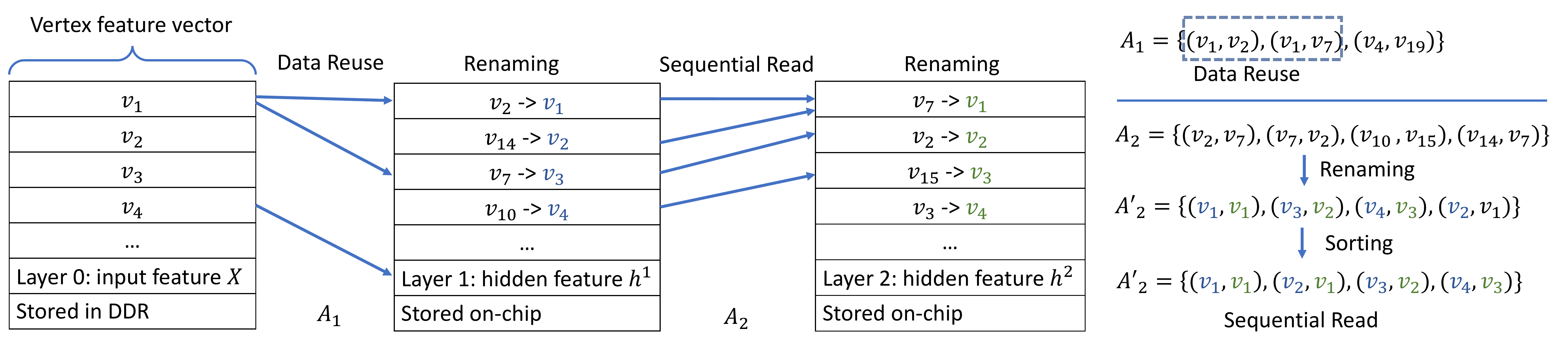}
    \vspace{-0.2cm}
    \caption{Data layout and Internal Representation}
     \label{fig:data}
     \vspace{-0.2cm}
\end{figure*}

\section{Accelerator Design}\label{HLS_template}
\begin{figure}[h]
    \centering
    \includegraphics[width=8.5cm]{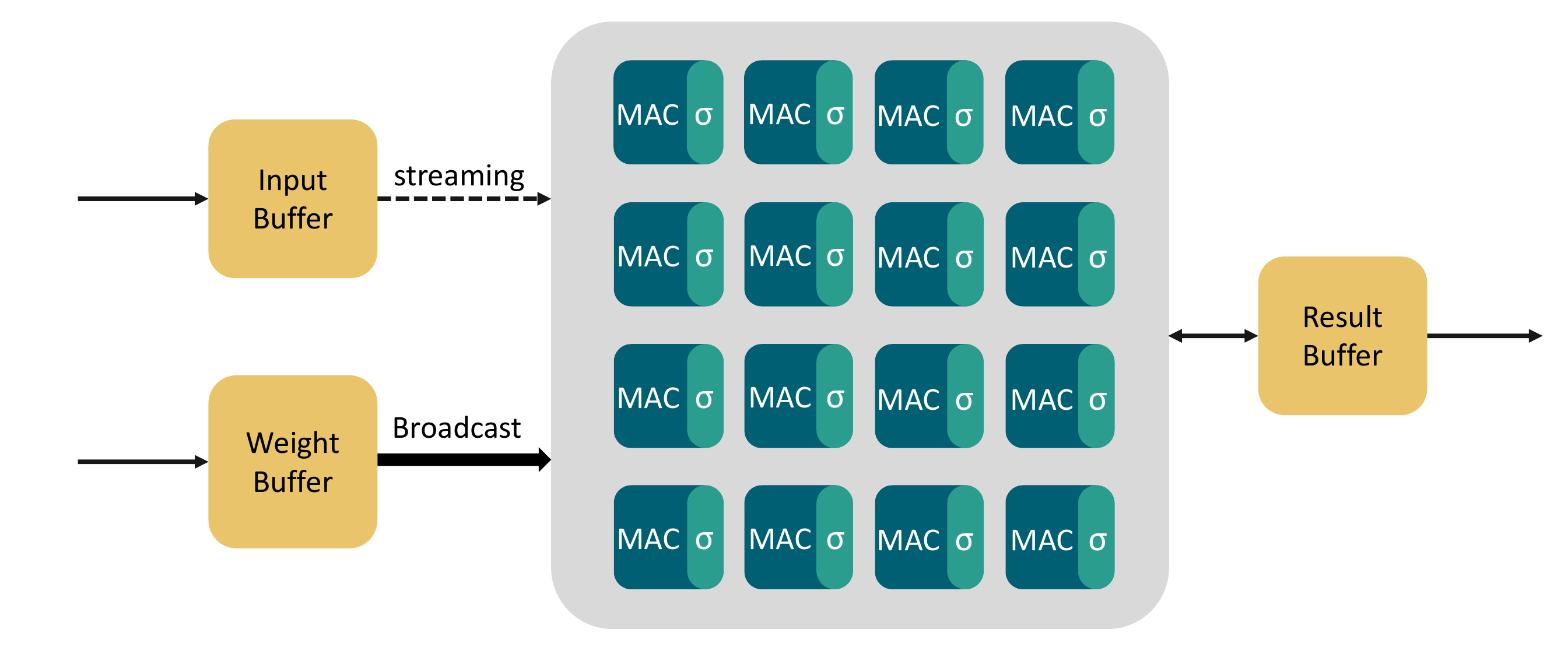}
    \vspace{-0.3cm}
    \caption{Architecture of update kernel}
    \vspace{-0.3cm}
     \label{fig:update}
\end{figure}

We design hardware templates based on the computation abstraction of the GNN layer described in Section \ref{GNN_train}. The hardware templates describe a general architecture of the GNN layer as in Figure \ref{fig:SPMM} and Figure \ref{fig:update}. Then, the accelerator generator takes user-defined functions as input and integrates them into the hardware templates to generate the accelerator design.

\subsection{Data Layout and Internal Representation}
\label{subsec:data-layout}
Aggregating feature vector from neighbors incurs  irregular memory access and large memory traffic. Figure \ref{fig:data} presents the data layout and internal representation used in our framework to reduce the memory traffic as well as the number of random memory accesses. The data layout is produced by the sampler through \emph{renaming} and \emph{sorting}.

\begin{algorithm}
\caption{Aggregation by Scatter-Gather Paradigm}\label{alg:scatter-gather}
\begin{algorithmic}

\While {not done}
\State {scatter phase:}
\For{each edge $e$ }
\If {edge.src = feat.src}
\State {Produce update $u \gets ${Scatter\_func($feat.val, edge.val$)}}
\EndIf
\EndFor

\State {gather phase:}
\For{each update $u$ }
\State {target vertex $v = v_{u.dst}$}
\State{Update vertex $v \gets$ {Gather\_func($u.val, v.val$)}}
\EndFor

\EndWhile
\end{algorithmic}
\end{algorithm}

\noindent\textbf{Reducing Memory Traffic (RMT)}: During aggregation stage, the feature vector of the source vertex is sent to its destination for aggregation. 
For the first layer of aggregation, the input feature matrix $\bm{X}$ is stored in the memory. Thus, loading feature vectors incurs a large number of random memory accesses.
Since the edges are represented in coordinate (COO) format sorted by source vertices in our framework, edges that share the same source vertex can reuse the feature vector that has been loaded, and thus reduce memory traffic. The total number of memory traffic can be reduced from  $O(|\mathcal{E}_{1}|f^{0})$ to $O(|\mathcal{B}_{0}|f^{0})$, where $|\mathcal{E}_{1}|$ is usually larger than $|\mathcal{B}_{0}|$. Fig \ref{fig:data} depicts an example: <$v_1,v_2$> loads the feature vector of $v_1$ from memory, and the loaded feature vector of $v_1$ can be reused by ($v_1,v_7$).

\noindent\textbf{Reducing Random Access (RRA)}: 
Since the edges are sorted by the source index in the first layer, the destination vertex index of the edges are in a random order; thus, the hidden features are stored randomly as shown in the layer 1 and layer 2 of Figure \ref{fig:data}. To reduce random access, our framework performs vertex renaming, which labels the vertices based on the order they are stored, this step also renames the vertices in each edge. Next, we sort the renamed edges by source vertices, and then accessing hidden features become sequential since the source vertex number follows the order it is stored.


\subsection{Kernel Design}\label{subsec:spmm}

 \begin{figure}[h]
    \centering
    \includegraphics[width=8.5cm]{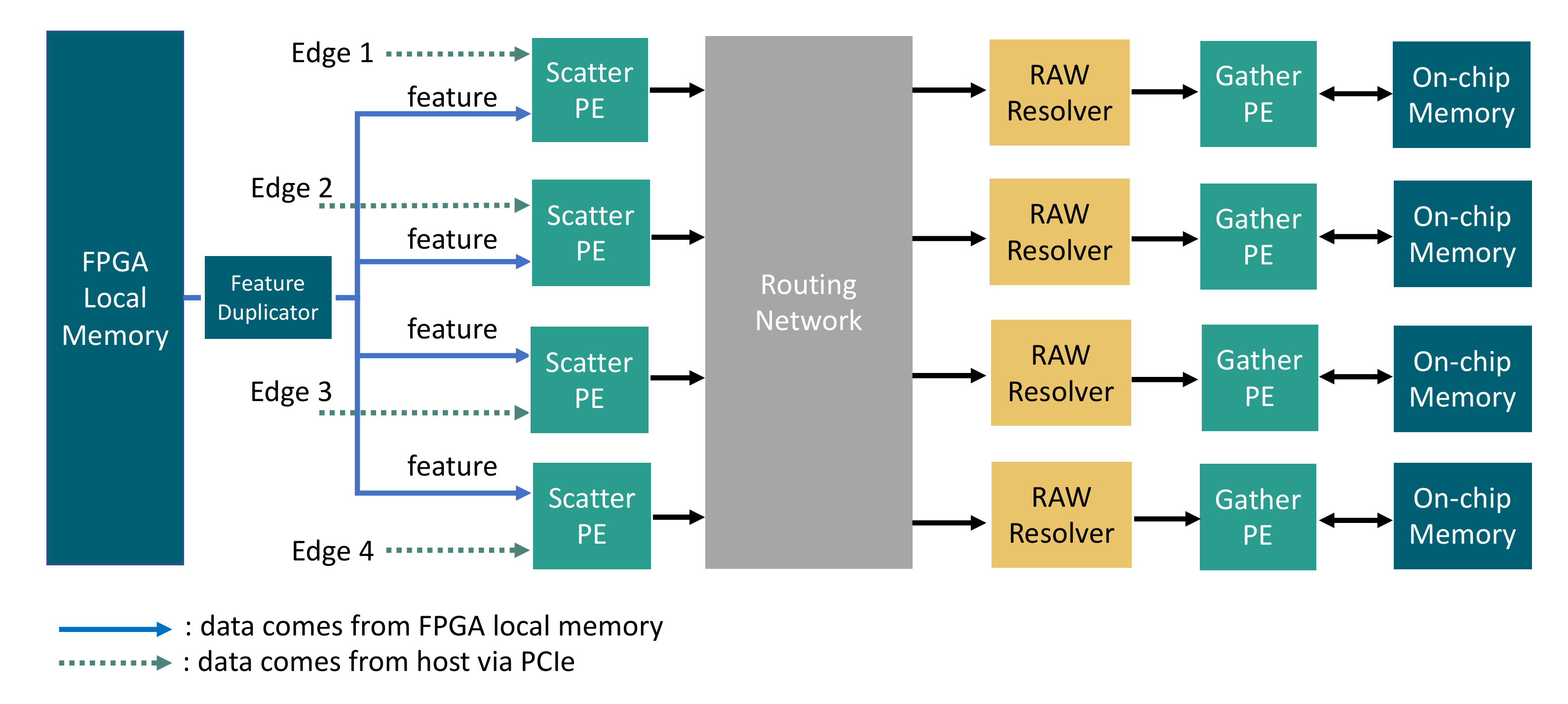}
    \vspace{-0.4cm}
    \caption{Architecture of aggregate kernel}
    \vspace{-0.3cm}
     \label{fig:SPMM}
\end{figure} 

\noindent \textbf{Aggregate Kernel}: The aggregate kernel adopts the scatter-gather paradigm \cite{thunderGP} as illustrated in Algorithm \ref{alg:scatter-gather}.
Figure \ref{fig:SPMM} depicts the detailed architecture of the aggregate kernel.
Multiple processing elements (PEs) process multiple edges concurrently in each clock cycle. 
The vertex feature vectors are first streamed to a feature duplicator. The feature duplicator broadcasts the loaded feature vector to all the Scatter PEs. The feature vector is stored in the PEs' registers for data reuse. Then, Scatter PEs  perform user-defined scatter() function, and stream the update $u$ to its destination via the routing network. The routing network is implemented as a butterfly network \cite{choi2021hbm}. After the Gather PEs receive the updates, Gather PEs perform user-defined gather function and obtain the intermediate results. The intermediate results are stored on-chip.  Finally, when the aggregation is done, the results stored in the on-chip memory are written back to the FPGA local memory.  Since the gather phase may incur reading and writing to the same destination vertices, read-after-write (RAW) data hazard may occur.
The RAW Resolver addresses RAW data hazard by stalling.


\noindent \textbf{Update Kernel}: The update kernel is a systolic array based design that performs block matrix multiplication. The input buffer loads the aggregation results $\bm{a}^l$ (see Algorithm \ref{alg:GNN}) from the FPGA local memory. $\bm{a}^l$ will then be streamed into the MAC array. Each MAC module is followed by an element-wise operator $\sigma$. Typically, weight of each layer $\bm{W}^l$ in GCN is small and frequently reused. Thus, $\bm{W}^l$ are stored on-chip in the Weight-Buffer. $\bm{W}^l$ will be broadcast to the multiply-accumulate (MAC) array during feature update. Finally, the result is stored into a result buffer before written back to the FPGA local memory.

\begin{figure}[h]
    \centering
    \includegraphics[width=6cm]{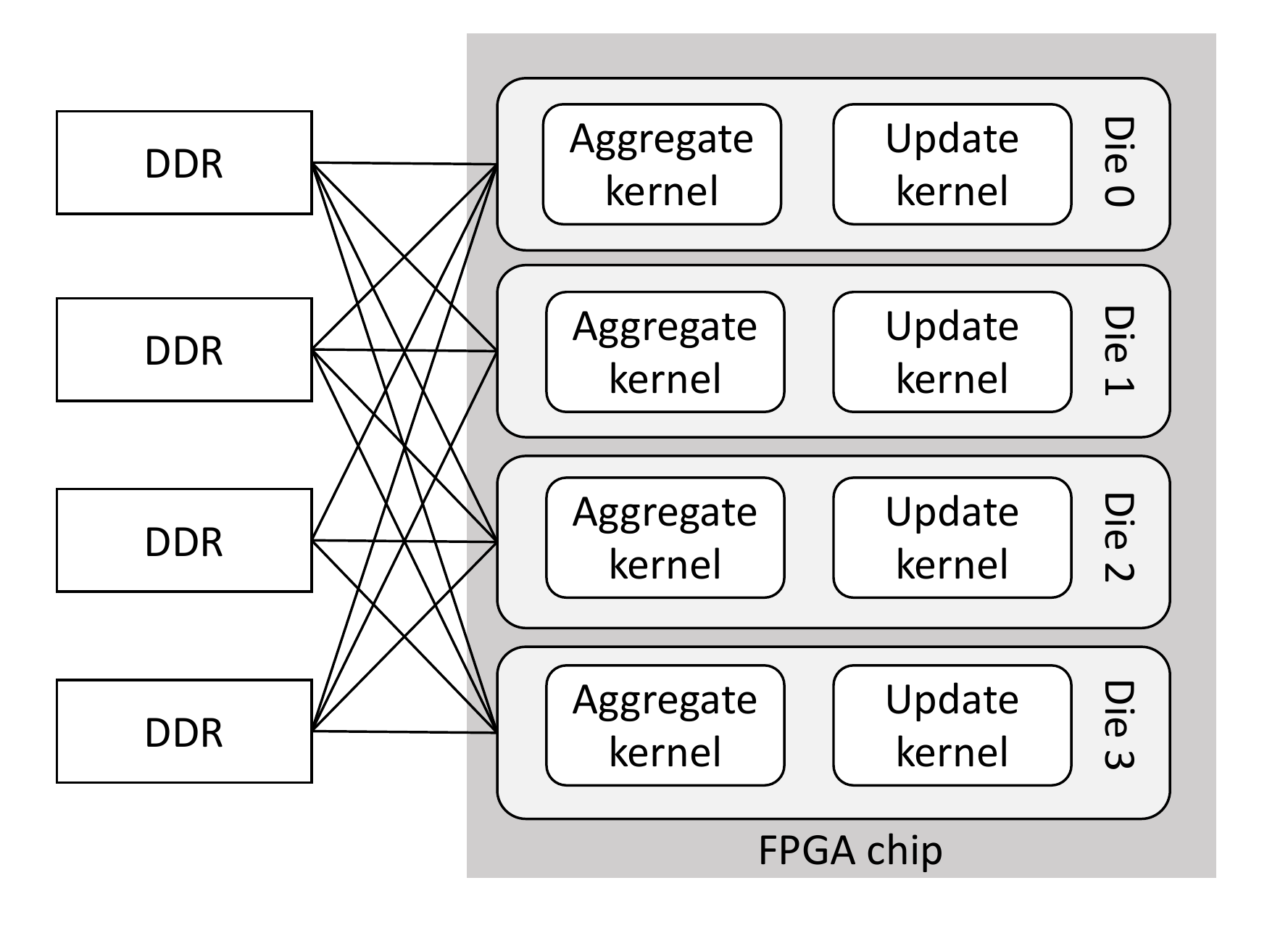}
    \vspace{-0.3cm}
    \caption{Architecture of the FPGA accelerator.}
     \vspace{-0.3cm}
     \label{fig:multi-dies}
\end{figure} 

\subsection{Parallel Computation Kernels}

Many modern FPGAs consists of multiple dies and number of interconnection wires across the dies is limited. Therefore, we implement multiple copy of the kernels that is distributed into multiple dies as shown in Figure \ref{fig:multi-dies}. Multiple dies and multiple DDR channels are connected through an all-to-all  interconnection which is generated by vendor tool, such as Xilinx Vitis. The input feature matrix $\bm{X}$ is equally partitioned into DDR channels. To utilize the multiple computation kernels for a single mini-batch, we perform task partitioning for the mini-batch training. In the forward propagation phase of layer 1, to infer the vertices in $\mathcal{B}^{1}=\{{v_{1},v_{2}, v_{3}, ...., v_{|\mathcal{B}^{1}|}}\}$. The workload for inferring $\mathcal{B}^{1}$ are equally partitioned into multiple kernels. Suppose there is $4$ aggregate kernels and $4$ corresponding update kernels. Aggregate kernel 1 aggregates $\{ {v_{1},v_{2}, ..., v_{\frac{|\mathcal{B}^{1}|}{4}} } \} $. Update kernel 1 updates $\{ {v_{1},v_{2}, ..., v_{\frac{|\mathcal{B}^{1}|}{4}} } \} $ and write the results back to DDR. Similarly, aggregate kernel 2 aggregates $\{ {{v_{\frac{|\mathcal{B}^{1}|}{4}} + 1},{ v_{\frac{|\mathcal{B}^{1}|}{4}} + 2}, ..., v_{\frac{|\mathcal{B}^{1}|}{2}} } \} $, and so on and so forth.   The same partitioning scheme is applied to each layer.

\section{Design Space Exploration}\label{DSE}
\label{sec:dse}
Our framework provides a design space exploration (DSE) engine for optimizing the GNN training throughput, given the configuration of mini-batch ($\{|\mathcal{B}^{l}|:0 \leqslant l \leqslant L\}$, $\{|\mathcal{E}^{l}|:1 \leqslant l \leqslant L\}$), GNN hidden dimensions $\{f^{l}:0 \leqslant l \leqslant L\}$, memory bandwidth $\alpha$, and hardware resources per die (DSPs, BRAMs, URAMs). To drive the optimization, we develop a performance model (Section \ref{subsec:performance-model}) that models the training throughput on the CPU-FPGA platform, and the resource utilization model (Section \ref{subsec:Resource-Utilization-Model}) that is used to specify the resource constraints. Then, our DSE engine (Algorithm \ref{alg:DSE}) performs parameter sweep in the design space to identify the hardware design parameters that optimizes the training throughput.



\subsection{Performance Model}
\label{subsec:performance-model}
We define the throughput of mini-batch GNN training as Number of Vertices Traversed Per Second (NVTPS): 

 \begin{equation}
   \text{Throughput} = \frac{\sum_{l=0}^L|\mathcal{B}^l|}{t_{\text{execution}}}
    \label{eq:throughput}
\end{equation}


The numerator indicates the total amount of vertices traversed in one mini-batch, and the denominator $t_{\text{execution}}$ is the average execution time of one training iteration (see Algorithm \ref{alg:GNN}).  The modeling of the average execution time is based on the our task scheduling on the CPU-FPGA platform.

We overlap sampling stage of the next batch with the execution of the current batch, so average execution time $t_{\text{execution}}$ is estimated as:
\begin{equation}
\begin{split}
    t_{\text{execution}} = \max \left(t_{\text{sampling}}, t_{\text{GNN}}\right)\\
    t_{\text{GNN}} = t_{\text{FP}} + t_{\text{LC}} + t_{\text{BP}} + t_{\text{WU}}
\end{split}
\end{equation}
where $t_{\text{\text{GNN}}}$ consists of the execution time of forward propagation $t_{\text{FP}}$, loss calculation $t_{\text{LC}}$, back propagation $t_{\text{BP}}$ and weight update $t_{\text{WU}}$.  


\noindent \textbf{Modeling $t_{\text{GNN}}$}: Loss calculation and weight update are executed on the host processor, which have optimized implementation in the software library.
Forward propagation and backward propagation are executed on the FPGA platform, and their execution time depends on the hardware parameters and the mini-batch configuration ($\{|\mathcal{B}^{l}|:0 \leqslant l \leqslant L\}$, $\{|\mathcal{E}^{l}|:1 \leqslant l \leqslant L\}$). We drive the approximate execution time of two propagation stages as:
\begin{equation}
    \begin{split}
        t_{\text{FP}} = \sum_{l=1}^L max(t_{\text{aggregate}}^l,t_{\text{update}}^l) \\
        t_{\text{BP}} = t_{\text{update}}^{1} + \sum_{l=2}^L max(t_{\text{aggregate}}^l,t_{\text{update}}^l) 
    \end{split}
\end{equation}
 The total propagation time $t_{\text{FP}}$ or $t_{\text{BP}}$ is the sum of the execution time of each layer, and the execution time of each layer is decided by the task that takes longer to complete since aggregation stage and update stage are pipelined.

%
The aggregation stage consists of two tasks: (1) loading vertex features or gradients, and (2) computation. Since the two tasks are pipelined, $t_{\text{aggregate}}$ can be modeled as:
 \begin{equation}
    t_{\text{aggregate}}^l = max(t_{\text{load}}^l,t_{\text{compute}}^l)
    \label{eq:agg}
\end{equation}

 \begin{equation}
 \begin{split}
      t_{\text{load}}^l = \frac{|\mathcal{B}^{l-1}|\times f^l \times S_{\text{feat}}}{BW\times\alpha} ~\text{ }~\text{ }~~~ t_{\text{compute}}^l = \frac{|\mathcal{E}^l|\times f^l }{n\times 16\times freq}
 \end{split}
    \label{eq:load}
\end{equation}

We model the vertex feature loading time $t_{\text{load}}^l$ as $\frac{\text{data transferred}}{\text{effective bandwidth}}$.
$|\mathcal{B}^l|$ indicates the number of vertices in each layer, $f^l$ is the feature length, and $S_{\text{feat}}$ is the data size of each feature. $\alpha$ is the effective bandwidth ratio. For the feature loading of the first layer of neighbor sampling method, $\alpha$ is estimated based on the memory burst transaction length $S_{\text{feat}}$ \cite{lu2021demystifying} (for DDR4) because it incurs random memory access; for the rest of the layers, $\alpha$ is near to $1$ \cite{lu2021demystifying} since the memory accesses are sequential from DDR memory or on-chip memory (Section \ref{subsec:data-layout}). The value of $\alpha$ is obtained from the prior work \cite{lu2021demystifying} which performs a profiling for the characteristics of the FPGA DDR memory.  We model the compute time as (\# of operations)/(\# of PEs $\times$ kernel frequency). $n$ denotes the there are $n$ Scatter PEs and $n$ Gather PEs instantiated in the aggregation kernel. $|\mathcal{E}^{l}|$ is the number of edges (i.e. non-zeros) in each layer. The size of $|\mathcal{E}^l|$ depends on the sampling method. We show the modeling of $\mathcal{E}^l$ for various sampling methods in Table \ref{tab:edge_size}. 
\begin{table}[!ht]
\centering
\caption{\# of vertices and \# of edges in each layer of various sampling methods}
\vspace{-0.2cm}
\begin{threeparttable}
\begin{adjustbox}{max width=0.45\textwidth}
\begin{tabular}{l|l|l}
 \toprule
  \textbf{Method} & \textbf{ \# of Vertices $|\mathcal{B}^l|$} & \textbf{\# of edges $|\mathcal{E}^l|$}   \\ 
\midrule
\midrule
\textbf{Neighbor}& $|\mathcal{V}^t|\times\Pi_{i=l+1}^L NS^i$ & $|\mathcal{V}^t|\times\Pi_{i=l}^L NS^i$
      \\ \midrule
\textbf{Layer-wise}& ${S}^l$ & ${S}^l\times{S}^{l-1}\times \kappa({S}^l)$    \\ \midrule
\textbf{Subgraph$^\ddagger$ }& $SB$ & $SB\times \kappa(SB) $    \\ \bottomrule
\end{tabular}
\end{adjustbox}
\begin{tablenotes}
\item[$\ddagger$] Uses node sampler in \cite{graphsaint-iclr20} as an example
\end{tablenotes}
\end{threeparttable}
\vspace{-0.2cm}
\label{tab:edge_size}
\end{table}
For neighbor sampling, the number of edges $|\mathcal{E}^{l}|$ in each layer is decided by the size of target vertices $|\mathcal{V}^t|$ and sample size $NS^l$. For layer-wise and subgraph sampling, we formulate the number of edges in layer $|\mathcal{E}^l|$ as
$|\mathcal{B}^l|\times|\mathcal{B}^{l-1}|\times\kappa(|\mathcal{B}^l|)$ where $|\mathcal{B}^l|\times|\mathcal{B}^{l-1}|$ corresponds to the case all the sampled vertices in layer $l$ and $l-1$ are connected, and $\kappa(|\mathcal{B}^l|)$ is a pre-trained function that estimates the graph sparsity based on sample size $|\mathcal{B}^l|$.

The feature update can be modelled as
\begin{equation}
    t_{\text{update}} = \frac{|B^l|\times f^l \times f^{l+1}}{m\times freq} 
    \label{eq:ns_update}
\end{equation}
 Similar to $t_{\text{compute}}$, we model the $t_{\text{update}}$ as (\# of operations)/(\# of PEs $\times$ kernel frequency). The numerator is the complexity of matrix multiplication, and $m$ denotes how many parallel MACs are instantiated in the update kernel.
 
 \noindent \textbf{Modeling  $t_{\text{sampling}}$}: the mini-batch sampling is performed on the host processor, which can potentially be a performance bottleneck. We exploit multi-threading to sample  multiple mini-batches concurrently. In the design phase, we estimate  $t_{\text{sampling}}$ under various number of threads and determine the minimum number of threads that satisfies $t_{\text{sampling}} < t_{\text{GNN}}$.

\subsection{Resource Utilization Model}
\label{subsec:Resource-Utilization-Model}

We set the hardware constraints to form the solution space for our DSE Engine. Among the various hardware resources on the FPGA platform, DSPs and LUTs are used the most as we increase the parallelism of the hardware modules. Thus, we model the usage of LUTs and DSPs as our constraints:
\begin{equation}
    \lambda_1\times m  + \lambda_2 \times n  \leq {N}_{DSP}
    \label{eq:DSP}
\end{equation}
\begin{equation}
    \rho_1\times m  + \rho_2 \times n  + \rho_3 \times n \log(n) \leq {N}_{LUT}
    \label{eq:LUT}
\end{equation}
The coefficients $\lambda_{i}~(1 \leqslant i \leqslant 2)$ and $\rho_{i}~(1 \leqslant i \leqslant 3)$ are constants that indicate the resource consumption for each PE. In the case of DSPs, the utilization grows linearly as we instantiate more PEs; In the case of LUTs, an additional $n\log(n)$ (see Section \ref{subsec:spmm} for the definition of $n$) term is introduced. The $n\log(n)$ LUT overhead models the routing network in the aggregation kernel shown in Figure \ref{fig:SPMM}. ${N}_{\text{DSP}}$ and ${N}_{\text{LUT}}$ denote the available DSPs and LUTs on the FPGA platform. 

\begin{algorithm}
\caption{Design Space Exploration Engine}
\label{alg:DSE}

\begin{algorithmic}[1]
\For{each die}
\State \textbf{Construct\_Search\_Space( )}
{\color{blue}\Comment{Derive $n_{max},m_{max}$}}
\For{$n=1...n_{max}$}
{\color{blue}\Comment{Exhaustive search}}
\For{$m=1...m_{max}$}
\State{{\color{blue}\#Check resource availability based on Eq. (\ref{eq:DSP}), (\ref{eq:LUT})}}
\State{Valid $\gets$ Check\_resource\_availability($n$,$m$)} 
\If{Valid \textbf{And} \textbf{Throughput(}$n$,$m$\textbf{)} $ > $ $max\_val$}
\State $max\_val$ $\gets$ \textbf{Throughput(}$n$,$m$\textbf{)}
\State \textbf{Save\_configuration(}$n$,$m$\textbf{)}
\EndIf
\EndFor
\EndFor
\EndFor
\end{algorithmic}
\end{algorithm}

\subsection{DSE Engine}
Many modern FPGAs consists of multiple dies \cite{chen2019cloud}, and the available resources may vary across dies. Thus, we perform DSE for each die to explore the optimal hardware configuration. We assume that each die is connected to  one DDR channel (e.g. Xilinx Alveo U250) for simplicity in Algorithm \ref{alg:DSE};
The DSE engine first constructs the search space by deriving the maximum value of $n$ and $m$ separately based on Equations (\ref{eq:DSP}) and (\ref{eq:LUT}). Then, the engine performs an exhaustive search through all the possible configurations. For each configuration, the engine evaluates its throughput using Equation \ref{eq:throughput}, and chooses the optimal design.

\section{Experiments}

\subsection{Experimental Setup}
We use our framework to generate GNN training implementations on a CPU-FPGA heterogeneous platform, and compare the training throughput with CPU-only platform and CPU-GPU platform. We list the information of each platform in Table \ref{tab:power}. The CPU-only and CPU-GPU baseline are implemented using PyTorch-Geometric\cite{Fey/Lenssen/2019} \footnote{\url{https://github.com/pyg-team/pytorch_geometric/blob/master/examples/reddit.py}} and GraphSAINT\cite{graphsaint-iclr20} \footnote{\url{https://github.com/GraphSAINT/GraphSAINT}}.

\begin{table}[!ht]
\centering
\caption{Specifications of the platforms }
\vspace{-0.2cm}
\begin{threeparttable}

\begin{adjustbox}{max width=0.47\textwidth}
\begin{tabular}{c|ccc}
 \toprule
\textbf{Platforms} & \begin{tabular}[|c|]{@{}c@{}} CPU \\  AMD Ryzen 3990x \end{tabular}  & \begin{tabular}[|c|]{@{}c@{}} GPU \\  Nvidia A100 \end{tabular} & \begin{tabular}[|c|]{@{}c@{}} FPGA \\  Xilinx Alveo U250 \end{tabular}  \\ 
\midrule \midrule
 {Technology}  & TSMC 7 nm   & TSMC 7 nm & TSMC 16 nm \\ 
{Frequency} & 2.90 GHz  & 1410 MHz & 300 MHz 
      \\ 
{Peak Performance}& 3.7 TFLOPS & 19.5 TFLOPS & 0.6 TFLOPS  \\ 
{On-chip Memory}& 256 MB L3 cache & 40 MB L2 Cache & 54 MB  \\
{Memory Bandwidth}& 107 GB/s & 1555 GB/s & 77 GB/s   \\ \bottomrule
\end{tabular}
\end{adjustbox}
\end{threeparttable}
\label{tab:power}
\end{table}




\noindent \textbf{Samplers, Models and Datasets}:
 We generate mini-batch for GNN training using two sampling algorithms: (1) GraphSAGE neighbor sampler \cite{hamilton2017inductive} for neighbor sampling (NS) and (2) GraphSAINT node sampler \cite{graphsaint-iclr20} for subgraph sampling (SS). For GraphSAGE neighbor sampler, we set the size of target vertices $|\mathcal{V}^{t}|$ as 1024, neighbor sampling size $NS$ as 25 and 10 for 1-hop neighbors and 2-hop neighbors; for GraphSAINT-node sampler, we set the sampling budget $SB$ as 2750.  We measure the GNN training throughput of two-layer GCN model and two-layer GraphSAGE model on four medium-scale graph datasets (Flickr \cite{graphsaint-iclr20}, Reddit \cite{hamilton2017inductive}, Yelp \cite{graphsaint-iclr20} and AmazonProducts \cite{graphsaint-iclr20}) that fit in the FPGA local DDR memory.  Details of the datasets and the GNN-layer dimensions are shown in Table \ref{tab: graph-scale}.
\begin{small}
\begin{table}[h]
\caption{Statistics of the Datasets and GNN-layer dimensions}
\vspace{-0.2cm}
    \centering
    \begin{adjustbox}{max width=0.47\textwidth}
    \begin{tabular}{ccccc}
        \toprule
        \textbf{Dataset} & \textbf{\#Nodes} & \textbf{\#Edges} & \hspace{-0.01cm} $f_{0}\hspace{0.4cm}  f_{1} \hspace{0.23cm} \hspace{0.03cm}f_{2}$\\
        \midrule
        \midrule
        Flickr (FL) & 89250  & 899756  & \hspace{-0.05cm}$500 \hspace{0.2cm} 256\hspace{0.25cm} 7$\\
        Reddit (RD) & 232965  & 11606919  & $ \hspace{0.05cm }602 \hspace{0.2cm} 256\hspace{0.2cm} 41 $\\
        Yelp (YP) & 716847 & 6977410 &  \hspace{0.05cm} $300\hspace{0.2cm} 256 \hspace{0.15cm} 100$\\
        AmazonProducts (AP) & 1598960 & 132169734 & \hspace{0.05cm} $200\hspace{0.2cm} 256 \hspace{0.15cm} 107$\\
        \bottomrule
    \end{tabular}
    \end{adjustbox}
    \label{tab: graph-scale}
\end{table}
\end{small}

\subsection{Framework Implementation}
\label{framework_imp}
We implement the program parser, DSE engine, software and hardware generator in our framework using Python, and the accelerator templates are implemented using Xilinx HLS. The host program template is programmed in OpenCL. Users interface with our framework using our APIs programmed in Python. To serve application developers, our framework includes several commonly used GNN models that can be used off-the-shelf. For ML researchers, our APIs allow users to define new models. In Listing \ref{lst:model}, we provide some examples of user inputs to specify platforms, mini-batch samplers and GNN models via our APIs. Based on the given inputs, our framework generates the host program and synthesizable accelerator design. For example, based on the GNN model specified by the user, user decides the parameters in the host program template, and what aggregation function should be filled in the HLS template to generate the accelerator design. Based on the platform parameters, our DSE engine fills in the hardware configurations such as the unroll factor in our HLS template. In Listing \ref{lst:HLS}, we provide an example that shows part of generated host program and synthesizable accelerator design. 

\begin{table}[h]
\caption{Resource utilization and Parallelism}
\vspace{-0.2cm}
    \centering
    \begin{adjustbox}{max width=0.46\textwidth}
    \begin{tabular}{c|cccc}
        \toprule
        \textbf{Resources} & NS-GCN & NS-GraphSAGE & SS-GCN & SS-GraphSAGE  \\
        \midrule \midrule
        LUTs  & 50\%  & 54\% & 44\%  & 76\%\\
        DSPs & 70\% & 54\% & 70\% & 82\% \\
        URAM &  34\% & 34\% & 14\% & 20\% \\
        BRAM & 28\%   & 28\% & 30\% & 34\%\\\midrule
        (m,n) & (256,4) &  (256,4) &  (256,4) &  (256,8)   \\
        \bottomrule
    \end{tabular}
    \end{adjustbox}
    \label{tab: resource}
\end{table}

In Table \ref{tab: resource} we show the resource utilization of the implementations generated by our design. The number $n$ which dentoes the number of Scatter PEs and Gather PEs are restricted to power of 2, and number of MACs $m$ is restricted to square of power of 2 due to the design of our accelerator.

%

\begin{lstlisting}[language=Python, caption=Examples of user inputs , label={lst:model}]
'''Example of platform specification'''
Parameter = PlatformParameters(board='xilinx-U250')
Parameter = PlatformParameters(board= , SLR = 4, DSP=3072, LUT=423000, URAM=320, BRAM= , BW=19.25)      #specify parameters for new boards

'''Example of different sampler'''
Samp = Sampler( 'NeighborSampler', L=2, budgets=[10,25])
Samp = Sampler( 'SubgraphSampler', L=2, budgets=[1500])

'''Example of different GNN models'''
GNN_para = GNN_Parameters(L=2, hidden=[256], v_feat)
GNN_comp = GNN_Computation('SAGE')
Model = GNN_Model(GNN_comp, GNN_para)

GNN_para = GNN_Parameters(L=2, hidden=[256], v_feat)
GNN_comp = GNN_Computation('customize')
Model = GNN_Model(GNN_comp, GNN_para)

'''User defined functions for custom layer operator'''
Scatter(edge, feat, msg){msg.dst = edge.dst; 
msg.val = edge.val*feat[edge.src];}

Gather(msg, v_ft){v_ft[msg.dst] += msg.val;}

Update('ReLU')
\end{lstlisting}

\vspace{-0.2cm}

\begin{lstlisting}[language=C, caption=Example of generated code, label={lst:HLS}]
//part of the generated host program
cl::Device devices = xcl::get_xil_devices();
for(int iter = 0; iter < Layer; iter++) {
// variable "Layer" is specified by user
    buffer_out =  cl::Buffer(CL_MEM_WRITE_ONLY);
    aggregare_krnls.setArg(0, buffer_out);
    ...//set arguments for kernel
    q.enqueueTask(aggregare_krnls);
    q.finish();
    q.enqueueMigrateMemObjects(buffer_out);
    //copy result back to host}
//part of the generated accelerator design
read_from_stream(input, tmp_update);
if(tmpupdate.valid == 1){
	index_type dst = tmp_update.dst - dst_offset;
	reg_update = result_buffer[dst];
	
	for (int j = 0; j < 16; j++){
	#pragma HLS unroll = 8 //decide by DSE
		regupdate.data[j] = regupdate.data[j] + tmpupdate.value.data[j];
		}//user-specified aggregate function
		
		resultbuffer[dst] = regupdate;
	}


\end{lstlisting}

\subsection{Impact of Optimizations}
We evaluate the two optimizations of our data layout and internal representation described in Section \ref{subsec:data-layout} on a two-layer neighbor sampling GCN. The two optimizations are: (1) reducing memory traffic (RMT) by reusing loaded vertex features in different edges that share the same source vertex; (2) reducing random access (RRA) by vertex renaming followed by edge sorting.  We first measure the throughput of the baseline implementation with no optimizations, and then incrementally apply the two optimizations. As shown in Table \ref{tab:improve}, both optimizations increase the GNN training throughput and can deliver up to 57\% improvement in total.





\begin{small}
\begin{table}[!ht]
\centering
\caption{Throughput improvement due to the optimizations}
\vspace{-0.2cm}
\begin{threeparttable}

\begin{adjustbox}{max width=0.46\textwidth}
\begin{tabular}{l|llll}
 \toprule
{Throughput (NVTPS)} & FL & RD & YP & AP   \\ 
\midrule
\midrule
{Baseline} &10.45 M  & 12.98 M & 19.71 M & 23.17 M 
      \\ 
{RMT}& 11.98 M & 16.48 M & 22.39 M & 27.22 M   \\ 
{RMT+RRA}& 16.38 M & 18.50 M & 24.60 M & 29.27 M  \\ \bottomrule
\textbf{Improvement}& \textbf{57\%} & \textbf{43\%} & \textbf{25\%} & \textbf{26\%}  \\ \bottomrule
\end{tabular}
\end{adjustbox}
\end{threeparttable}
\label{tab:improve}
\end{table}
\end{small}
\vspace{-0.2cm}

\begin{small}

\begin{table}[]
\centering
\caption{Cross Platform Comparison (Throughput)}
\vspace{-0.2cm}
\begin{threeparttable}
\begin{adjustbox}{max width=0.47\textwidth}
\begin{tabular}{lllll}
 \toprule
                         & \textbf{Data} & \textbf{CPU}       & 
                         \textbf{CPU-GPU}        & \textbf{CPU-FPGA}      \\ \midrule \midrule
\multirow{4}{*}{NS-GCN}  & FL      & 265.5K (1$\times$) & 2.69M (10.1$\times$)  & 16.38M (61.7$\times$)  \\ \cline{2-5} 
                         & RD      & 85.65K (1$\times$) & 7.15M (83.5$\times$)  & 18.50M (216$\times$)   \\ \cline{2-5} 
                         & YP      & 275.6K (1$\times$) & 9.36M (34.0$\times$)  & 24.61M (89.2$\times$)  \\ \cline{2-5} 
                         & AM      & 480.6K (1$\times$) & 13.0M (29.0$\times$)   & 29.26M (60.8$\times$)  \\ \midrule
\multirow{4}{*}{NS-SAGE} & FL      & 225.2K (1$\times$) & 2.74M (12.2$\times$)  & 11.84M (52.6$\times$)  \\ \cline{2-5} 
                         & RD      & 78.50K (1$\times$) & 6.90M (88.0$\times$)  & 13.10M (166$\times$)   \\ \cline{2-5} 
                         & YP      & 266.0K (1$\times$) & 9.19M (34.5$\times$)  & 18.12M (68.1$\times$)  \\ \cline{2-5} 
                         & AM      & 479.3K (1$\times$) & 13.57M (28.3$\times$)   & 21.15M (44.1$\times$)  \\ \midrule
\multirow{4}{*}{SS-GCN}  & FL      & 215.2K (1$\times$) & 768.3K (3.59$\times$)   & 2.81M (13.0$\times$) \\ \cline{2-5} 
                         & RD      & 118.9K (1$\times$) & 536.4K (4.51$\times$)   & 2.56M (21.5$\times$) \\ \cline{2-5} 
                         & YP      & 159.1K (1$\times$) & 751.0K (4.71$\times$)   & 3.08M (19.4$\times$) \\ \cline{2-5} 
                         & AM      & 25.55K (1$\times$) & OoM            & 1.47M (57.5$\times$) \\ \midrule
\multirow{4}{*}{SS-SAGE} & FL      & 179.9K (1$\times$) & 626.7K (3.48$\times$)   & 2.71M (15.1$\times$) \\ \cline{2-5} 
                         & RD      & 94.72K (1$\times$) & 505.2K (5.33$\times$)   & 2.43M (25.6$\times$) \\ \cline{2-5} 
                         & YP      & 126.7K (1$\times$) & 709.7K (5.60$\times$)   & 2.78M (22.0$\times$) \\ \cline{2-5} 
                         & AM      & 17.40K (1$\times$) & OoM            & 1.45M (83.3$\times$) \\ \midrule
\textbf{Average}                      &         & \textbf{193.4K (1$\times$)} & \textbf{4.96M (25.66$\times$)} & \textbf{10.77M (55.67$\times$)} \\
 \bottomrule
\end{tabular}
\vspace{-0.2cm}
\end{adjustbox}
\end{threeparttable}
\label{tab:perf}
\end{table}

\end{small}

\subsection{Comparison with State-of-the-art}
For evaluation, we use the throughput defined in Section \ref{DSE} as metric, i.e. number of vertices traversed per second (NVTPS).
 To measure the throughput, we count the total number of vertices in each mini-batch, and measure the average execution time of one training iteration. Table \ref{tab:perf} shows the throughput comparison of GNN training across the three platforms. All three implementations use single precision floating point as data type.

Comparing with the CPU-only baseline, the CPU-GPU platform achieves $25.66\times$ throughput on the average. This is because CPU-GPU platform provides massive data parallelism with $5.27\times$ peak performance and $14.5\times$ memory bandwidth compared with the CPU-only platform (Table \ref{tab:power}). 
Comparing with the CPU-only baseline and CPU-GPU baseline, the CPU-FPGA implementation generated by our framework achieves $55.67\times$ and $2.17\times$ throughput on the average respectively. Though CPU-GPU platform has higher memory bandwidth and peak performance than CPU-FPGA platform, the throughput is limited by the memory access overhead during aggregation stage. While our accelerator can access the on-chip memory in one cycle (3.3 ns), CPU and GPU need to access the data in multi-level caches. Taking AMD Ryzen 3990 as an example, the L2 cache latency is 5 to 12 ns, and the L3 cache latency is around 32 ns.
Moreover, as shown in Table \ref{tab:improve}, our data layout and internal representation also improves the training throughput by reducing the memory traffic and reducing random memory accesses. 

We compare our results with two state-of-the-art GNN training implementations: GraphACT \cite{10.1145/3373087.3375312} and Rubik \cite{9428002}. As shown in Table \ref{tab:perf-cmp-SOTA}, our framework achieves up to $4.45\times$ and $3.4\times$ throughput respectively. Compared with GraphACT, the achieved speedup is due to (1) the vertex features are fetched directly from the FPGA local memory, (2) the proposed aggregate kernal has higher computation parallelism compared with Feature Aggregation Module in GraphACT. Compared with ASIC design Rubik, the obtained speedup is due to (1) larger on-chip memory of FPGA that can fully store the intermediate results under the setting of the experiments, (2) our proposed data layout optimizations that reduce the external memory traffic and random memory accesses.  

\begin{small}
\begin{table}[]
\centering
\caption{Comparison with State-of-the-art}
\vspace{-0.2cm}
\begin{threeparttable}
\begin{adjustbox}{max width=0.47\textwidth}
\begin{tabular}{ccccc}
 \toprule
                         &      &  \textbf{GraphACT} \cite{10.1145/3373087.3375312}$\ddagger$      & \textbf{Rubik}  \cite{9428002}        & \textbf{This work}      \\ \midrule  \midrule
 \multirow{4}{*}{\begin{tabular}[|c|]{@{}c@{}} \textbf{Platform}  \end{tabular}} & \textbf{Device} & Alveo U250 & ASIC & Alveo U250 \\ \cline{2-5} 
   &  \textbf{Peak  Perf.}  & 0.6 TFLOPS & 1 TFLOPS  & 0.6 TFLOPS  \\ \cline{2-5} 
    & \textbf{Bandwidth} & 77 GB/s &   432 GB/s &    77 GB/s             \\ \cline{2-5} 
    & \textbf{On-chip Mem.} & 54 MB & 2 MB &   54 MB             \\  
                    \midrule 
\multirow{2}{*}{\begin{tabular}[|c|]{@{}c@{}} \textbf{SS-SAGE} \\ (Throughput) \end{tabular}} & \textbf{RD} & 546.8K (1$\times$) & 717.0K (1.31$\times$) & 2.43M (4.45$\times$) \\ \cline{2-5} 
                         & \textbf{YP}   & 769.8K (1$\times$) &     N/A              & 2.78M (3.61$\times$) \\
 \bottomrule
\end{tabular}
\end{adjustbox}
\begin{tablenotes}
\item[$\ddagger$] Scaled from U200 to U250 using the number of DSPs.
\end{tablenotes}
\vspace{-0.2cm}
\end{threeparttable}

\label{tab:perf-cmp-SOTA}

\end{table}
\end{small}




\section{Discussion}
We discuss the novelty of this work compared with previous work GraphACT \cite{10.1145/3373087.3375312} and the applicability of proposed optimizations to other platforms (e.g., CPU, GPU).

\vspace{0.1cm}
\emph{Comparison with GraphACT}. In GraphACT \cite{10.1145/3373087.3375312}, the redundancy reduction requires that all the edges have uniform weight value. Therefore, it can not support GCN \cite{gcn}. In constrast, the proposed optimizations such as RMT and RRA do not have requirements on the weight, therefore, can support broader range of GNN models. Moreover, the Feature Aggregation Module in GraphACT has limited computation parallelism in feature-level that limits its performance for neighbor-sampling-based GNN training. In comparison, the proposed aggregate kernel adopts the scatter-gather paradigm with routing network, which can enable massive computation parallelism within feature aggregation.

\emph{Optimizations on CPU/GPU platforms}. In this work, we proposed a highly optimized aggregate kernel that adopts the scatter-gather paradigm to accelerate feature aggregation. While the scatter phase is optimized by our proposed data layout optimizaiton (Section \ref{subsec:data-layout}), the performance of gather phase depends on the routing network of  aggregate kernel  (Section \ref{subsec:spmm}) that efficiently routes the intermediate result from Scatter PEs to Gather PEs. On CPU/GPU platforms, the data layout optimization can potentially be adopted to optimize the scatter phase. However, the gather phase is hard to be optimized on CPU/GPU since the data communication among the computation cores  is through a complex cache hierarchy. Since the scatter phase and gather phase need be optimized  simultaneously, the proposed optimizations may lead to limited performance improvement on CPU/GPU platforms.

\section{Conclusion}

In this paper, we proposed HP-GNN, a general framework to generate high-throughput  GNN training implementation on a given CPU-FPGA heterogeneous platform. Based on the high-level abstraction of GNN computation, we designed a host program template and hardware templates to support various GNN models. Our proposed data layout and internal representation improve the throughput of GNN training. The implementations generated by HP-GNN achieve $55.67\times$ and $2.17\times$ throughput compared with state-of-the-art CPU-only and CPU-GPU platforms. Compared with state-of-art accelerators, our framework achieves up to $4.45\times$ throughput. In the future, we plan to extend our framework to multi-FPGA platforms by exploiting model parallelism. 


\begin{acks}
We would like to thank the anonymous reviewers for their comments which greatly improved the presentation. This work has been supported by the U.S. National Science Foundation  under  grant  numbers  OAC-1911229, CNS-2009057 and  CCF-1919289.  Equipment  and  support  by  Xilinx  are  greatly  appreciated.
\end{acks}


\balance
\bibliographystyle{ACM-Reference-Format}
\bibliography{refer}

\end{document}